\documentclass[11pt]{article}
\usepackage[latin1]{inputenc}
\usepackage[english]{babel}
\usepackage[namelimits]{amsmath}
\usepackage{amssymb}
\usepackage{amsmath}
\usepackage{amsthm}

\begin{document}
\title{Rigidly rotating dust solutions depending upon  
harmonic functions}
\author{Stefano Viaggiu
\\
Dipartimento di Matematica,\\
Universit\'a di Roma ``Tor Vergata'',\\
Via della Ricerca Scientifica, 1, I-00133 Roma, Italy\\
E-mail: viaggiu@mat.uniroma2.it\\
(or: stefano.viaggiu@ax0rm1.roma1.infn.it)}
\date{\today}\maketitle
\begin{abstract}
We write down the relevant
field equations for a stationary axially symmetric rigidly rotating 
dust source
in such a way that the general solution depends upon
the solution of an elliptic equation and upon
harmonic functions. Starting with the dipole Bonnor solution,
we built an asymptotically flat solution with two curvature singularities
on the rotational axis 
with diverging mass.
Apart from the two
point singularities on the axis, 
the metric is regular everywhere. Finally, we study
a non-asymptotically flat solution with NUT charge and a massless ring 
singularity, but with a well-defined mass-energy expression.
\end{abstract}

\section*{Introduction}
The problem of building a physically admissible metric for an
isolated rotating body is still long an unresolved problem
\cite{1,9}. In fact, in
order to obtain a physically reasonable source, many restrictions must
be imposed (energy conditions, regularity, reasonable equation of state).
In particular, Einstein's equations for a rotating body with
a perfect fluid source 
seem not to be integrable. A remarkable exception is given
by dust pressureless stationary axially symmetric spacetimes. In a remarkable
paper \cite{10}, Winicour showed that Einstein's equations for a
stationary axially symmetric dust source 
with differential rotation can be reduced
to quadratures. These equations contain as a subclass the van Stockum one
\cite{11} of rigidly rotating matter. The first asymptotically flat
solution of the van Stockum class can be found in \cite{12}. The author in
\cite{12} shows that the van Stockum class of solutions that are not 
cylindrically symmetric cannot exsist in the Newtonian theory.
The solution in \cite{12} has a curvature singularity 
with diverging mass. Further, the 
technique named ''displace, cut, reflect'' \cite{13} to obtain rotating discs
immersed in rotating dust must be noted. Unfortunately, this method
generates distributional exotic matter on the $z=0$ plane.\\
In this paper, starting from the Lewis \cite{14,15} form of the metric,
we write down the equations for stationary axially symmetric rigidly rotating
spacetimes in a co-moving reference frame in such a way that
the general solution depends upon the solution of an elliptic equation and
upon harmonic functions. The class of solutions contains the van
Stockum line element for a suitable choice of the harmonic function.\\
In this context, starting from the dipole Bonnor solution \cite{12}, we 
obtain an asymptotically flat solution with two curvature singularities 
on the rotation axis and
showing simalar properties to the Bonnor solution.\\ 
In section 1 we derive the basic equations. In section 2 we present our
solution. Section 3 collects some final remarks and conclusions. In
the appendix we derive a non asymtotically flat solution with
a NUT charge and a well-defined mass-energy expression. 

\section{Basic Equations}
Our starting point is the Lewis \cite{14} line element for a stationary 
axisymmetric space-time:
\begin{equation}
ds^2=e^{v(\rho,z)}\left[d{\rho}^2+dz^2\right]+L(\rho,z)d{\phi}^2+
2 m(\rho,z)dtd\phi -f(\rho,z)dt^2,
\label{1}
\end{equation}
where $x^4=t$ is the time coordinate, $x^1=\rho$ is the radial coordinate
in a cylindrical system, $x^2=z$ is the zenithal coordinate and $x^3=\phi$
is the azimuthal angular coordinate on the plane $z=0$. Also,
\begin{equation}
t \in (-\infty,\infty)\;,\;\rho \in (0,\infty)\;,\;z \in (-\infty,\infty)\;,\;
\phi\in [0,2\pi).
\label{1b}
\end{equation}
Further, the root square of the
determinant of the 2-metric spanned by the 
Killing vectors ${\partial}_t, {\partial}_{\phi}$ is
\begin{equation}
\sqrt{|det\;g^{(2)}|}=\sqrt{fL+m^2}=W(\rho,z).
\label{2}
\end{equation}
Expression (\ref{2}) characterizes the measure of the area of the orbits 
of the isometry group. In the vacuum, the field equations for (\ref{1}) 
imply that $W(\rho,z)$ is harmonic, i.e. $W_{,\alpha,\alpha}=0$, 
where subindices denote
partial derivative and a summation with respect to  $\alpha=\rho,z$
is implicit. Therefore, the function $W(\rho,z)$ can be chosen as a
coordinate.
Looking for regular
solutions on the axis, the simplest assumption  can be made by setting  
$W=\rho$. In this way, the van Stockum line element emerges by 
taking a dust source. However, this is not the
most general choice. Thanks to the gauge freedom, we can take
\begin{equation}
fL+m^2={\rho}^2 H(\rho,z),
\label{4}
\end{equation}
where $H(\rho,z)$ is a sufficiently regular function to be specified by the
field equations.\\
We consider a perfect fluid 
$T_{\mu\nu}=(E+P)u_{\mu}u_{\nu}+Pg_{\mu\nu}$, with $E$ being the mass-energy
density , $P$ the hydrostatic pressure and $u_{\mu}$ the 4-velocity
of the fluid.       
We consider a co-moving reference frame:
\begin{equation}
u^t=\frac{1}{\sqrt{f}}\;\;,\;\;u^{\phi}=u^{\rho}=u^{z}=0.
\label{6}
\end{equation}
Denoting with $R_{\mu\nu}$ the Ricci tensor,
the relevant field equations are
\begin{eqnarray}
& &R_{zz}-R_{\rho\rho}=0, \label{8}\\
& &R_{\rho\rho}+R_{zz}=(P-E)e^{v}, \label{9}\\
& &R_{\rho z}=0, \label{10}\\
& &R_{\phi\phi}=\left[\frac{L}{2}T-T_{\phi\phi}\right],\label{11}\\
& &R_{t\phi}=\left[\frac{m}{2}T-T_{t\phi}\right],\label{12}\\
& &R_{tt}=-\left[\frac{f}{2}T+T_{tt}\right],\label{13}
\end{eqnarray}
where $T=3P-E$. Equation (\ref{11}) involves a second-order partial equation
for $L(\rho,z)$, while (\ref{12}) and (\ref{13}) give second order equations
for $m(\rho,z)$ and $f(\rho,z)$ respectively. Thanks to (\ref{4}), equations
(\ref{11})-(\ref{13}) are not independent. Therefore, from (\ref{4}), we
can express $L(\rho,z)$ in terms of $(f,m,H)$. Putting this expression in
(\ref{11}) and using equations (\ref{12}) and (\ref{13}), we obtain
the following compatibility equation:
\begin{equation}
4HPe^v=H_{,\alpha,\alpha}-\frac{H^{2}_{,\alpha}}{2H}+\frac{2}{\rho}H_{,\rho}.
\label{14}
\end{equation}
In what follows we study dust solutions for which $P=0$. Setting
${\rho}^2 H=F^2$, equation (\ref{14}) becomes 
$F_{,\alpha,\alpha}=0$. Thus the compatibility condition for
(\ref{11})-(\ref{13}) requires that $F(\rho,z)$ be a harmonic function. 
Conversely, with $F(\rho,z)$ no more harmonic, 
the line element (\ref{1}) is 
appropriate to describe spacetimes with non-vanishing pressure $P$.
For $F=\rho$ the van Stockum line element is regained together with the
Papapetrou form of the metric \cite{15}. Equations (\ref{8}) and 
(\ref{10}) are linear first-order equations involving $v_{,\rho}$ and
$v_{,z}$ and they permit us to calculate $v_{,\rho}, v_{,z}$ in terms of
$(f,m,F)$. By applying the integrability condition 
($v_{,\rho,z}=v_{,z,\rho}$) for the equations so obtained, we read
\begin{equation}
f_{,z}F_{,\rho}=F_{,z}f_{,\rho}.
\label{15}
\end{equation} 
Finally, when expressions for $v_{,\rho}, v_{,z}$ are put in (\ref{9}),
we obtain
\begin{equation}
F_{,z}f_{,z}=-f_{,\rho}F_{,\rho}.
\label{16}
\end{equation}
Excluding the case $F=const$ (it can be see that this leads to the trivial
solution $E=0$), we have $f=const$. We naturally choose $f=1$.
Therefore our system of equations is
\begin{eqnarray}
& &F_{,\alpha,\alpha}=0, \label{17}\\
& &m_{,\alpha,\alpha}-\frac{m_{,\alpha}F_{,\alpha}}{F}=0,\label{18}\\
& &E=\frac{e^{-v}m^{2}_{,\alpha}}{F^2},\label{19}\\
& &v_{,\rho}=\frac{\left[m^{2}_{z}F_{,\rho}-
m^{2}_{,\rho}F_{,\rho}+4FF_{,z}F_{,z,\rho}-4FF_{,\rho}F_{,z,z}-
2m_{,\rho}m_{,z}F_{,z}\right]}{2FF^{2}_{,\alpha}},\label{20}\\
& &v_{,z}=\frac{\left[4FF_{,z}F_{,z,z}+4FF_{,\rho}F_{,z,\rho}
-F_{,z}m^{2}_{,z}+F_{,z}m^{2}_{,\rho}-2m_{,\rho}m_{,z}F_{\rho}\right]}
{2FF^{2}_{,\alpha}},\label{21}\\
& &L+m^2=F^2.\label{22}
\end{eqnarray}
First of all, for non-expanding spacetimes,
the shear $q_{ik}=\frac{1}{2}[u_{i;k}+u_{k;i}]$ vanishes
identically for (\ref{17})-(\ref{22}), and therefore our system of equations
describes rigidly rotating sources in a co-moving
reference frame. When $F=\rho$, 
equation (\ref{18}) is invariant under the transformation
$z\rightarrow z+a$ ($a$ a constant) and a solution can be expanded
as $\sum_{i}m(\rho, z+a_i)$. Setting $F\neq\rho$, if $F(\rho,z), m(\rho,z)$
are solutions, then also $F(\rho,z+a), m(\rho,z+a)$ are, and thus
the solutions cannot be expanded.\\
Note that we have identified $(\rho,z)$ with the radial 
and the zenithal coordinate respectively in a cylindrical coordinate system. 
According to this assumption, some conditions must be imposed.
Firstly, by setting $E=0$, the metric must reduce to the standard flat
expression $ds^2=d{\rho}^2+dz^2+{\rho}^2d{\phi}^2-dt^2$. Therefore, 
$\lim_{E \to 0}F=\rho$. Further, looking for regular spacetimes on the
rotation axis , the norm of the space-like Killing vector 
${\partial}_{\phi}$ must be vanishing (except at isolated points)
at $\rho =0$, i.e.
$\lim_{\rho \to 0}L=0$. Finally, for asymptotically flat spacetimes, at
spatial infinity $F(\rho,z)$ looks as follows:
$F=\rho +o(1)$. 

\section{Generating an asymptotically flat solution}
Our starting point is the Bonnor dipole solution \cite{12} 
$F=\rho , m=\frac{c{\rho}^2}{{({\rho}^2+z^2)}^{\frac{3}{2}}}$. We
can obtain a solution of (\ref{18}) by taking the map 
$\rho\rightarrow F(\rho,z), z\rightarrow G(\rho,z)$, where 
$F=\rho\left(1+\frac{bc}{({\rho}^2+z^2)}\right) ,
    G=z\left(1-\frac{bc}{({\rho}^2+z^2)}\right)$ with $c\geq 0$, being 
 $b$ a constant.
Therefore we get the solution
\begin{eqnarray}
& &F=\rho\left(1+\frac{bc}{({\rho}^2+z^2)}\right),\label{B1}\\
& &m=
\frac{c{\rho}^2{[{\rho}^2+z^2+bc]}^2}
     {\sqrt{{\rho}^2+z^2}{[{({\rho}^2+z^2)}^2+2bc{\rho}^2+b^2c^2-
     2bcz^2]}^{\frac{3}{2}}},\nonumber\\
& &v=\ln(\alpha)+\frac{c^2{\rho}^2}{8}
\frac{\gamma{({\rho}^2+z^2+bc)}^2}
     {{[{({\rho}^2+z^2)}^2+2{\rho}^2 bc+b^2c^2-2z^2 bc]}^{4}},\nonumber\\
& &E=\frac{c^2 e^{-v}\beta\Delta}
     {{[{({\rho}^2+z^2)}^2+2{\rho}^2 bc+b^2c^2-2z^2 bc]}^{4}},\nonumber\\
& &\beta = \alpha {({\rho}^2+z^2)}^2=
        {({\rho}^2+z^2)}^2+c^2b^2-2{\rho}^2 cb+2z^2 cb,\nonumber\\
& &\gamma = 
({\rho}^2-8z^2){({\rho}^2+z^2)}^2+2{\rho}^4 cb+18{\rho}^2z^2 bc+
{\rho}^2 c^2b^2+16z^4 bc-8z^2 c^2b^2,\nonumber\\
& &\Delta =
({\rho}^2+4z^2){({\rho}^2+z^2)}^2-8z^4 cb+4z^2c^2b^2-
6{\rho}^2 z^2 cb+{\rho}^2 c^2b^2+2{\rho}^4 cb.\nonumber   
\end{eqnarray}
Solution (\ref{B1}) is asymptotically flat. Note that the map
$\rho\rightarrow F(\rho,z)\;,\;z\rightarrow G(\rho,z)$ is not bijective,
i.e. is not a diffeomorphism.
Concerning the features of (\ref{B1}), they depend on the sign of
the constant $b$.\\
For $b>0$, apart from $\rho =0, z=\pm\sqrt{bc}$, our solution is regular
everywhere. At these two points, we have
curvature singularities with properties close to the $\rho =0, z=0$
singularity of the dipole Bonnor solution (see \cite{12}). In
particular, the mass-energy diverges at these points.
Otherwise, the energy density $E(\rho,z)$ is integrable.
For $b=0$, we regain the Bonnor solution.
Finally, for $b<0$, the two point singularities disappear, but
emerges a curvature ring singularity for $z=0, \rho=\sqrt{|b|c}$
with diverging mass. Independently on the parameter $b$,
at spatial infinity the metric reduces
to the standard expression in 
asymptotical cylindrical coordinates, and so also by setting $c=0$ ($E=0$).
Note that, because of the non-invertibility of the map between the Bonnor
solution and solution (\ref{B1}), the curvature singularity at the 
origin of \cite{12} is shifted in the two curvature singularities of
(\ref{B1}) (for $b>0$) on the rotation axis.   
As a final consideration, it must be noted that there exists 
for (\ref{B1}) a finite
non singular region about the origin. 
Thus, our solution could be matched, in principle, with some 
asymptotically flat vacuum solution. We do not enter in this
discussion, but only mention this possibility.  

\section{Conclusions and final remarks}
We have studied stationary axially symmetric rigidly
rotating dust spacetimes in terms of harmonic functions.
In \cite{B} Bonnor found the general solution for charged dust with zero 
Lorentz force in terms of harmonic functions. However, the use of 
such kind of functions in \cite{B} is different from the one in our paper.
In fact, in our paper harmonic functions appear in equation (\ref{4})
thanks to the gauge freedom, while in \cite{B} the function $F(\rho,z)$
is chosen to be the cylindrical polar coordinate $\rho$. 
In the Bonnor paper, harmonic functions arise
in order to obtain the most general solution for equation (\ref{18})
with $F=\rho$. In this case, all the solutions of the equation
$m_{,\alpha,\alpha}-\frac{1}{\rho}m_{,\rho}=0$ are given by taking
a generic harmonic function $\eta(\rho,z)$, with $m=\rho{\eta}_{\rho}$.
In a similar way, another harmonic function is introduced when charged
dust comes in action. Therefore, a direct relation between does not exist 
between
the harmonic function $F(\rho,z)$ and $\eta(\rho,z)$ of \cite{B}.\\
Also, the paper \cite{G} must be noticed in which charged dust solutions are 
given in terms of Bessel functions of first and second kind and hyperbolic
functions. Also in the paper \cite{G} the condition
$F=\rho$ is retained.\\
In this paper, section two, 
starting with the dipole Bonnor solution, we build a class
of asymptotically
flat solutions containing the Bonnor one as a subclass by a suitable
choice of the functions $F(\rho,z), G(\rho,z)$.
Obviously, it must be noted that not all the harmonic functions generate
physically sensible solutions. For a physically sensible solution we mean
a regular (apart from isolated singularities) asymptotically flat
solution. Generally, it is a simple matter to verify that, if
$F(\rho,z)=\rho, m(\rho,z)$ is a regular differentiable solution
for (\ref{18}), then also $m(F(\rho,z),G(\rho,z))$ is a solution 
for (\ref{18}) with
$F(\rho,z)$ harmonic being $G(\rho,z)$ the harmonic conjugate to $F(\rho,z)$
i.e. $F_{\rho}=G_z, F_z=-G_{\rho}$. Further, in order to generate a new  
asymptotically flat and regular solution on the axis starting with a seed
solution with these two properties, we must build a non-bijective
(not a diffeomorphism) map $\rho\rightarrow F(\rho,z), 
z\rightarrow G(\rho,z)$ such that
$\lim_{E \to 0}F=\rho\;,\;\lim_{E \to 0}G=z$, and
$\lim_{\rho \to 0}L=\lim_{\rho \to 0}F=0$
(apart from isolated points)
and such that
at spatial infinity the functions $F(\rho,z), G(\rho,z)$ look as follows:
$F=\rho+o(1)\;,\;G=z+o(1)$. 
Concerning isolated singularities, no general conclusions can be made.
The functions $F(\rho,z), G(\rho,z)$ of section two satisfy all the
conditions mentioned above.\\
Finally, in the appendix we present a non-asymptotically flat solution
not obtained from a seed solution with the technique discussed above and 
therefore it represents an ad hoc solution.
In particular, it is possible to build ad hoc solutions 
for (\ref{18}) by setting
$F(\rho,z)=\rho(1+\frac{c}{{\rho}^2+z^2})$ (with $c$ a constant)
and  $m(\rho,z)$ a homogeneous function such that 
$m_{,\alpha,\alpha}-\frac{m_{,\rho}}{\rho}=0$.
\section*{Appendix}
We consider the following solution:
\begin{eqnarray}
& &F=\rho\left[1+\frac{c}{{\rho}^2+z^2}\right]\;\;,\;\;
m=c\left[\frac{z}{\sqrt{{\rho}^2+z^2}}-1\right],\label{24}\\
& &v=\left[1+\frac{c}{16}\right]
  \ln[{\rho}^4-2c{\rho}^2+2{\rho}^2z^2+c^2+z^4+2cz^2]\nonumber\\
&  &-\frac{c}{8}\ln[c+{\rho}^2+z^2]-
     2\ln[{\rho}^2+z^2]\;,\;
E=\frac{c^2 e^{-v}}{{[c+{\rho}^2+z^2]}^2},\nonumber
\end{eqnarray}
with $c$ a constant. When $c>0$, the 
solution (\ref{24}) has a curvature ring 
singularity when $(\rho,z)=(\sqrt{c},0)$ 
The axis is regular
for $z >0$, 
while it shows a conical (no curvature) singularity when $z\leq 0$.
Further, for $z<0$ there is a region where 
closed time-like curves (CTC) appear resulting in a vioaltion of causality.
However, it is possible to take a simple 
coordinate transformation found in \cite{20}, i.e.
$\tau =t+2c\phi$
giving the whole rotational
axis free of conical singularities.
Unfortunately, we are forced to introduce a periodic time coordinate $\tau$
and therefore once again CTC appear. Therefore, the problem of violation
of causality cannot be avoided with an opportune coordinate transformation.\\
The spacetime asymptotically reads the expression appropriate for 
asymptotic NUT metrics \cite{16,17} with NUT charge $q$ given by
$c=2q$.
We can estimate the mass 
inside an infinite cylinder of radius $R$ by means of the integral
\begin{equation}
M(R)=\int_{-\infty}^{\infty}dz \int_{0}^{R}d\rho
\int_{0}^{2\pi}E Fe^{v}d\phi.
\label{29}
\end{equation}
The integral (\ref{29}) is well defined everywhere.
Thus, for the mass we get the formula
\begin{equation}
M(R)=2{\pi}^2 c^2\left[1+R-\sqrt{R^2+1}\right].
\label{30}
\end{equation}
Because of the non-asymptotical flatness,
the solution (\ref{24}) is not interesting in an astrophysical
context.
However, the natural arena for this solution is in the extra
relativistic context given by non-Abelian gauge theories or in the low 
energy string theory \cite{22,23} where, in order to obtain supersymmetries,
NUT charge comes in action.

\end{document}